\documentclass[aps,twocolumn,floatfix,nofootinbib,showpacs,superscriptaddress]{revtex4-1}

\usepackage{float,amsmath,amsfonts,amssymb,slashed,bm,graphicx,color}

\begin{document}

\title{Quark Spectral Function and Deconfinement at Nonzero Temperature}
\author{$\text{Si-xue Qin}$}
\email{sixueqin@th.physik.uni-frankfurt.de}
\affiliation{Institute for Theoretical Physics, Johann Wolfgang Goethe
  University, Max-von-Laue-Str.\ 1, D-60438 Frankfurt am Main, Germany}
\author{$\text{Dirk H.\ Rischke}$}
\email{drischke@th.physik.uni-frankfurt.de}
\affiliation{Institute for Theoretical Physics, Johann Wolfgang Goethe
  University, Max-von-Laue-Str.\ 1, D-60438 Frankfurt am Main, Germany}
\affiliation{Frankfurt Institute for Advanced Studies, 
Ruth-Moufang-Str.\ 1, D-60438 Frankfurt am Main, Germany}

\begin{abstract}
The  maximum entropy method is used to compute the quark spectral
function at nonzero temperature. We solve the gap equation of
quantum chromodynamics (QCD) self-consistently, employing a rainbow kernel 
which phenomenologically models
results from Dyson-Schwinger equations (DSE) and lattice QCD. 
We use the criterion of positivity restoration of the spectral 
function as a signal for
deconfinement. Our calculation indicates that the critical temperature
of deconfinement $T_d$ is slightly smaller than the one of chiral 
symmetry restoration $T_c$: $T_d\sim 94\% T_c$ in the chiral limit,
and $T_d\sim 96\% T_c$ with physical light quark masses. Since these deviations
are within the systematic error of our approach, it is reasonable to
conclude that chiral symmetry restoration and deconfinement coincide at
zero chemical potential.
\end{abstract}

\date{\today}

\maketitle

\section{Introduction}
Heavy-ion experiments at the Relativistic Heavy-Ion Collider (RHIC) 
and the Large Hadron Collider (LHC) are focusing on charting the 
phase diagram of hot and dense nuclear matter. The quark-gluon plasma
(QGP), a primordial state of matter in the early Universe, where
chiral symmetry is restored and quarks and gluons are deconfined, 
has been re-created in the extremely hot environment of a heavy-ion
collision. With the expansion of the fireball, nuclear matter 
cools down and dilutes. The low-temperature, low-density phase
of nuclear matter is characterized by confinement and dynamical
chiral symmetry breaking (DCSB). It is a central goal of modern 
theoretical physics to understand the properties of, and the transitions
between, these phases. The chiral and deconfinement 
phase transitions as well as their interplay are especially
interesting. In general, the chiral condensate (or, equivalently, 
the dynamical quark mass) is adopted as an order parameter for the 
chiral phase transition. The order of this transition may depend
on the number of quark flavors, the values of the quark masses, 
and whether the $U(1)_A$ anomaly of QCD is effectively restored.
For nonzero quark masses, the existence of a critical end point 
(where the transition
turns from being first order at low temperatures and high densities to
being crossover at high temperatures and low densities) has been
suggested but, even if it exists at all,  
its precise location is still highly debated. 

Concerning the
deconfinement phase transition, the situation is even more complicated
because confinement has been a mystery since the inception
of the Standard Model. The notion of confinement is easily understood
from the linearly rising potential between 
infinitely heavy quarks \cite{Wilson:1974sk,Eichten:1978dh}, which has
also been studied by lattice QCD \cite{Bali:2000gf}.
However, this is no longer true for light quarks because of strong
pair-creation and -annihilation effects \cite{Bali:2005bz}. 
In the pure-gauge limit realized for infinitely heavy
quarks, the center $Z(3)$ symmetry of the color gauge group $SU(3)$ is
preserved in the confining phase, while it is spontaneously broken in
the deconfined phase. Here, the Polyakov loop (or the thermal Wilson
line) \cite{Svetitsky:1982gs,Pisarski:2000eq} is the corresponding 
order parameter. Equivalently, the dual quark condensate 
\cite{Bilgici:2008qy,Fischer:2009wc,Fischer:2009gk} was proposed as an order
parameter, which makes it possible to study the interplay between 
confinement and DCSB. However, their validity as order parameters
for phase transitions in light-quark systems remains unclear. 

Besides these order parameters, confinement can be related to the 
analytic properties of QCD Schwinger functions 
\cite{ROBERTS:1992jd,Roberts:1994hs,ROBERTS:2008cq}. 
The axiom of reflection positivity requires that the propagator must 
have a positive definite K\"allen-Lehmann spectral representation 
for asymptotic (or deconfined) quarks. In other words, if the quark 
propagator can be decomposed in terms of 
complete eigenstates of the Hamiltonian, each of which should have 
positive probability, quarks can propagate as asymptotic states; 
otherwise, quarks have to be somehow confined. 
It can be shown that pairs of complex conjugate poles of the
full quark propagator lead to a nonpositive definite K\"allen-Lehmann spectral
representation. Therefore, in Refs.\
\cite{Maris:1995ns,Bhagwat:2002tx}, the existence of such pairs of
complex conjugate poles was considered
as a criterion for confinement. However, this is only a 
sufficient condition. The reason is that
the violation of reflection positivity can also be realised by
propagators with real poles, namely if they have a negative residue which
also leads to the K\"allen-Lehmann spectral representation 
being not positive definite \cite{Yuan:2010iy}. 
On the other hand, the positivity of the quark spectral function 
is a necessary and sufficient condition for quark deconfinement, 
no matter whether 
the singularities of the quark propagator are located on or off the real
axis. In short, by considering the quark spectral function directly, 
one is able to distinguish confined phases (where reflection positivity is
violated) from deconfined ones (where reflection positivity holds).

In this work, we use the maximum entropy method (MEM) 
\cite{Bryan:1990tva,Nickel:2007bm,Mueller:2010ah,Asakawa:2000tr} to explicitly 
compute quark spectral functions from the self-consistent
numerical solution of the QCD gap equation.
We employ a rainbow kernel \cite{Qin:2011kka,Qin:2011xq} which 
phenomenologically models recent results from DSE
\cite{Aguilar:2009ik,Aguilar:2012rz} and lattice QCD 
\cite{Bowman:2004gt,Oliveira:2011ds,Boucaud:2010gr}. We 
define a deconfinement temperature as the temperature
above which the positivity of the quark spectral function is
restored. 
This work is a continuation of Ref.\ \cite{Qin:2011hp} where the quark spectral
functions were studied in the region above $T_c$.
This paper is organized as follows. In Sec.\ \ref{sec:gapeq}, 
we present the QCD gap equation and discuss the \emph{ansatz}
employed for its solution. In Sec.\ \ref{sec:specrep}, we derive 
the relation between the quark spectral function 
and the solution of the gap equation. Here, we also define 
the order parameter which signals the positivity of the spectral
function. In Sec.\ \ref{sec:mem}, we briefly outline the MEM and its 
extension for nonpositive definitive spectral functions. 
Section \ref{sec:numrlt} reports our numerical results. Finally, we 
conclude with a summary and some remarks.

\section{QCD gap equation}\label{sec:gapeq}
At nonzero temperature, the QCD gap equation is
\begin{eqnarray}
\label{eq:gap1}
S(i\omega_n,\vec{p}\,)^{-1} &=&  i\vec{\gamma}\cdot\vec{p}
  + i\gamma_4 \omega_{n} + m
  + \Sigma(i\omega_n,\vec{p}\,) \, ,\\
\nonumber
\Sigma(i\omega_n,\vec{p}\,) &=& \frac{4T}{3}\sum_{l=-\infty}^{+\infty}
\! \int\frac{d^3\vec{q}}{(2\pi)^3}\; {g^{2}} D_{\mu\nu} 
(\vec{k}, \Omega_{nl})\\
& & \times {\gamma_{\mu}}\, S(i\omega_l,\vec{q}\,)\,
\Gamma_{\nu}(\vec{q}, \omega_{l},\vec{p},\omega_{n})\, ,
\label{eq:gap2}
\end{eqnarray}
where $\omega_n=(2n+1)\pi T$ is the fermionic Matsubara frequency, 
$\vec{k}=\vec{p}-\vec{q}\,$, $\Omega_{nl} = \omega_{n} - \omega_{l}$, 
$D_{\mu\nu}$ is the dressed gluon propagator, 
and $\Gamma_{\nu}$ is the dressed quark-gluon vertex.  
The solution of the gap equation can be expressed as
\begin{eqnarray}\label{eq:qdirac}
\nonumber
S(i\omega_n,\vec{p}\,)^{-1} & = & \,i\vec{\gamma} \cdot \vec{p}\, 
A(\omega_{n}^2,\vec{p}\,^2) \\
&& +\, i\gamma_{4} \omega_{n} C(\omega_{n}^2,\vec{p}\,^2) 
+ B(\omega_{n}^2,\vec{p}\,^2)
\end{eqnarray}
or, equivalently,
\begin{eqnarray}
\nonumber
S(i\omega_n,\vec{p}\,) & = & -\,i\vec{\gamma} \cdot \vec{p}\, 
\sigma_A(\omega_{n}^2,\vec{p}\,^2) \\
&& -\, i\gamma_{4} \omega_{n} \sigma_C(\omega_{n}^2,\vec{p}\,^2) 
+ \sigma_B(\omega_{n}^2,\vec{p}\,^2) \, ,\label{eq:quarkLz}
\end{eqnarray}
where $A,B,C$, and $\sigma_{A,B,C}$ are scalar functions. 
The dynamical quark mass is defined as 
$M(\omega_{n}^2,\vec{p}\,^2)=B(\omega_{n}^2,\vec{p}\,^2)/
A(\omega_{n}^2,\vec{p}\,^2)$, which is independent of the 
renormalization point. In the chiral limit, the chiral condensate 
is defined as
\begin{eqnarray}
\nonumber
-\langle \bar{q}q \rangle^0 & = & N_c T\sum_{n=-\infty}^{+\infty}
\int\frac{d^3\vec{p}}{(2\pi)^3} {\rm tr_D} S(i\omega_n,\vec{p}\,) 
\, ,\label{eq:cond}\\
&\sim& M(\omega_0^2,\vec{p}\,^2=0)\,.
\end{eqnarray}
However, because of an ultraviolet divergence
the integral in the above equation is not well defined at nonzero current quark mass. 
Conveniently, $M_0:=M(\omega_0^2,\vec{p}\,^2=0)$ can be used as 
the order parameter for the chiral phase transition, which is 
equivalent to the chiral condensate.

The gap equation is closed by specifying the vertex 
and the gluon propagator. Here, we use the 
rainbow truncation, i.e., the leading term in a symmetry-preserving 
scheme \cite{Bender:1996kg}:
\begin{eqnarray}
&&g^2 D_{\mu\nu}(\vec{k}, \Omega_{nl})\Gamma_\nu(\vec{q}, 
\omega_{l},\vec{p},\omega_{n}) \notag\\
&&= [P_{\mu\nu}^{T} D_{T}(\vec{k}\,^2, \Omega_{nl}^2) 
+ P_{\mu\nu}^{L} D_{L}(\vec{k}\,^2, \Omega_{nl}^2)]\gamma_\nu \,,
\end{eqnarray}
where $P_{\mu\nu}^{T,L}$ are transverse and longitudinal 
projection operators, respectively,
\begin{eqnarray}
P^T_{\mu\nu}&=&\left\{
\begin{aligned}
&0, \qquad \qquad \quad {\mu}\text{ and/or } {\nu} = 4 \, ,  \\
&\delta_{ij}-\frac{{\vec{k}_{i}} {\vec{k}_{j}}}{\vec{k}^2}, 
\quad {\mu}, {\nu} =1,2,3 \, ,
\end{aligned}
\right. \\
P^{L}_{\mu\nu} &= &\delta_{\mu\nu} - \frac{k_\Omega^\mu k_\Omega^\nu
}{k_\Omega^2} - P^T_{\mu \nu}\,,
\end{eqnarray}
with ${k_{\Omega}}:=(\Omega_{nl},\vec{k})$, and where
\begin{eqnarray}
D_{T} &=&\mathcal{D}(\vec{k}\,^2 + \Omega_{nl}^2)\, , \quad
D_{L} =\mathcal{D}(\vec{k}\,^2 + \Omega_{nl}^2 + m^2_g)\, .
\end{eqnarray}
Here, the function
\begin{eqnarray}
\mathcal{D}(s) & = &
\frac{8{\pi^{2}}D}{\sigma^4}e^{-s/\sigma^2} 
+ \frac{8{\pi^{2}} {\gamma_{m}}}{{\ln}[ \tau \! + \! (1 \! + \!
s/{\Lambda_{\text{QCD}}^{2}} ) ^{2} ] } \,{\cal F}(s)\,,\quad
\end{eqnarray}
with ${\cal F}(s) = [1-\exp(-s/4 m_t^2)]/s$, $\tau=e^2-1$, 
$m_t=0.5\,$GeV, $\gamma_m=12/25$, and 
$\Lambda^{N_f=4}_{\text{QCD}}=0.234$~GeV. 
For pseudoscalar and vector mesons with masses$\,\lesssim 1\,$GeV, 
this interaction provides a uniformly good description of 
their vacuum properties when $\sigma D=(0.8\,{\rm GeV})^3$ 
and $\sigma\in[0.4,0.6]\,{\rm GeV}$ \cite{Qin:2011kka,Qin:2011xq}, 
which means that there is only one free parameter in the model. 
The physical masses of the light quarks are $m_{u=d}^\zeta=3.4\,{\rm
  MeV}$ at our renormalization point $\zeta=19\,{\rm
  GeV}$. Generalizing 
to $T\neq 0$, we have followed perturbation theory and included a
Debye-like mass in the longitudinal part of the gluon propagator: 
$m_g^2= (16/5) T^2$ [for details, see Ref.\ \cite{Qin:2011hp}].

\section{Spectral representation}\label{sec:specrep}
The dressed quark propagator is related to the retarded
real-time propagator by analytic continuation,
\begin{eqnarray}
S^R(\omega,\vec{p}\,)=S(i\omega_n,\vec{p}\,)|_{i\omega_n\rightarrow\omega
+i\epsilon}.
\end{eqnarray}
From the spectral reprentation of $S^R(\omega,\vec{p}\,)$,
i.e.,
\begin{eqnarray}
\rho(\omega,\vec{p}\,) = -2\Im S^R(\omega,\vec{p}),
\end{eqnarray}
one immediately obtains
\begin{eqnarray}
S(i\omega_n,\vec{p}\,)=\int_{-\infty}^{+\infty}\frac{d\omega'}{2\pi}
\frac{\rho(\omega',\vec{p})}{i\omega_n-\omega'}\,.
\end{eqnarray}
According to Eq.\ \eqref{eq:quarkLz}, the spectral function can be 
decomposed as
\begin{eqnarray}\label{eq:rdirac}
\rho(\omega,\vec{p}\,)&=&-\,i\vec\gamma\cdot\vec{p}\,
\rho_v(\omega,\vec{p}\,^2)\notag\\
&&+\,\gamma_4\omega\,\rho_e(\omega,\vec{p}\,^2)+\,
\rho_s(\omega,\vec{p}\,^2)\,.
\end{eqnarray}
As a consequence of the anti-commutation relation, the spectral 
function fulfills the following sum rule,
\begin{eqnarray}
\int_{-\infty}^{+\infty}\frac{d\omega}{2\pi} \, \rho(\omega,\vec{p}\,)\gamma_4 
= \mathbf{1}\,.
\end{eqnarray}
Then one can define the spectral function 
$\rho_0(\omega,\vec{p}\,^2):=\omega\rho_e(\omega,\vec{p}\,^2)$, 
which is nonnegative
and can be treated as a probability distribution for deconfined quarks.
Note that $\rho_0(\omega,\vec{p}\,^2)$ can be easily related to 
the dressed quark propagator by
\begin{eqnarray}
S_0(\omega_n^2,\vec{p}\,^2)&=&i\omega_n\sigma_C(\omega_n^2,\vec{p}\,^2)\,,
\notag\\
&=&\int_{-\infty}^{+\infty}\frac{d\omega'}{2\pi}
\frac{\rho_0(\omega',\vec{p}\,^2)}{\omega'-i\omega_n}\,,\label{eq:specrep1}
\end{eqnarray}
where $\sigma_C$ is the scalar function in Eq.\ \eqref{eq:quarkLz}, 
or to the imaginary-time quark propagator
\begin{eqnarray}
D_0(\tau,\vec{p}\,^2)&=&T\sum_n e^{-i\omega_n\tau}
S_0(\omega_n^2,\vec{p}\,^2),\notag\\
&=&\int_{-\infty}^{+\infty}\frac{d\omega}{2\pi}\frac{e^{(1/2-\tau T)
\omega/T}}{e^{\omega/2T}+e^{-\omega/2T}}\,\rho_0(\omega,\vec{p}\,^2)\,.\quad
\label{eq:specrep2}
\end{eqnarray}

The above equations connect the quark spectral function which we
consider to the numerical solution of the gap equation. 
As a signal for positivity violation (or restoration) of 
the spectral function, one can define an ``order'' parameter as
\begin{eqnarray}
\hat{Z}_\rho&=&\int_{-\infty}^{+\infty}\frac{d\omega}{2\pi} |\rho(\omega)|\,,\\
Z_\rho&=&\int_{-\infty}^{+\infty}\frac{d\omega}{2\pi} \rho(\omega)\,, \\
L_\rho&=&\frac{\hat{Z}_\rho - Z_\rho}{\hat{Z}_\rho}\,,
\end{eqnarray}
where $\rho(\omega)$ simply denotes $\rho_0(\omega,\vec{p}\,^2=0)$. 
It is apparent that $\hat{Z}_\rho=Z_\rho=1$ and $L_\rho=0$ for a
positive definite spectral function, otherwise 
$\hat{Z}_\rho>Z_\rho=1$ and $L_\rho>0$.
The critical temperature $T_d$ of the deconfinement transition 
is defined as the lowest temperature where $L_\rho=0$.

\section{Maximum entropy method}\label{sec:mem}
It is an ill-posed problem to extract the spectral function from the
(imaginary-time) quark propagator. Actually, there is an infinite 
set of spectral functions which can reproduce a given correlation 
function with tolerable errors. The MEM 
\cite{Bryan:1990tva,Nickel:2007bm,Mueller:2010ah,Asakawa:2000tr} considers the 
probability distribution of spectral functions
to produce the most probable one. The theoretical basis 
is Bayes' probability theorem. The
conditional probability of having the spectral function 
$\rho(\omega)$ given the correlation function 
$D(\tau)$ reads 
\begin{eqnarray} 
P[\rho|DM]=\frac{P[D|\rho M]P[\rho|M]}{P[D|M]}\,,
\end{eqnarray} 
where $M$ summarizes all definitions and prior knowledge of the
spectral function, $P[D|\rho M]$ and $P[\rho|M]$ are called the
likelihood function and the prior probability, respectively.
Since $P[D|M]$ is independent of $\rho(\omega)$, it can be
treated as a normalization constant. 

According to the central-limit theorem the data $D(\tau)$ are 
expected to obey a Gaussian distribution:
\begin{eqnarray} 
P[D|\rho M]=\frac{1}{Z_L}e^{-L[\rho]}\,,
\end{eqnarray} 
with 
\begin{eqnarray} 
L[\rho]=\frac{1}{\beta}\int_0^{\beta} d\tau \frac{|D(\tau) 
- D[\rho](\tau)|^2}{2\xi(\tau)^2}\,,
\end{eqnarray} 
where $Z_L$ is a normalization constant, $D[\rho]$ denotes the 
correlation function reproduced by Eqs.\ \eqref{eq:specrep1} or 
\eqref{eq:specrep2} given the spectral function
$\rho(\omega)$, and $\xi(\tau)$ is the variance of the error. 
Maximizing the likelihood function is 
equivalent to $\chi^2$-fitting.

The construction of the prior probability $P[\rho|M]$ is the central 
idea of MEM, which expresses the prior in terms of the spectral entropy as
\begin{eqnarray} 
P[\rho|M(\alpha )]=\frac{1}{Z_S}e^{\alpha S[\rho,m]}\,,
\end{eqnarray} 
where $Z_S$ is a normalization constant and $\alpha$ is an undetermined 
positive scale factor. The Shannon-Jaynes entropy
$S$ is defined as 
\begin{eqnarray} 
S[\rho,m]=\int_{-\infty}^{+\infty}d\omega\left[\rho(\omega)-m(\omega)- 
    \rho(\omega)\text{ln}\frac{\rho(\omega)}{m(\omega)}\right]\,,\quad
\end{eqnarray} 
where $m(\omega)$ is the ``default model'' of the spectral function. 
Its typical form is a uniform distribution without
{\it a priori} structure assumption \cite{Qin:2011hp}, i.e.,
\begin{eqnarray} 
m(\omega)=m_{0}\theta(\Lambda^2-\omega^2)\,.
\end{eqnarray} 
Note that a reliable output from the MEM should be insensitive 
to $m_0$ and $\Lambda$.
If the spectral function is not positive definite, one can decompose it in terms of 
two positive definite components, i.e.,
\begin{eqnarray} 
\rho(\omega)=\rho_+(\omega)-\rho_-(\omega)\,,
\end{eqnarray} 
Correspondingly, the total entropy is expressed as 
\cite{Hobson:1998bz,Balandin:1999ab}
\begin{eqnarray} 
S[\rho,m]=S[\rho_+,m_+] + S[\rho_-,m_-]\,,
\end{eqnarray} 
where $m_\pm$ denotes the default models of $\rho_\pm$, respectively. 

Finally, one obtains the total probability distribution 
\begin{eqnarray} 
P[\rho|DM(\alpha )] \propto e^{\alpha S[\rho,m]-L[\rho]}\,.
\end{eqnarray} 
The most probable spectral function $\rho_\alpha(\omega)$ for fixed
$\alpha$ can be 
obtained by maximizing $P[\rho|DM(\alpha )]$, where usually
the standard 
singular-value decomposition algorithm of Bryan \cite{Bryan:1990tva} 
is adopted. To deal with the scale factor $\alpha$, we 
follow Bryan's Method \cite{Bryan:1990tva}. The MEM spectral function
is defined as
\begin{eqnarray} 
\rho_{\rm MEM} & = & \int_0^\infty d\alpha \int {\cal D} \rho\,
\rho(\omega) \,P[\rho|D M(\alpha)]\,P[\alpha|DM] \nonumber \\
&\simeq& \int_0^\infty d\alpha \,
\rho_\alpha(\omega)\,P[\alpha|DM]\,,\label{eq:specmem}
\end{eqnarray} 
where it is assumed that $P[\rho|DM(\alpha )]$ is sharply peaked 
around $\rho_\alpha(\omega)$, so that the functional integral over
$\rho$ can be approximated. In this way, the MEM
spectral function becomes an average of the $\rho_\alpha(\omega)$'s
with respect to $\alpha$. The conditional probability $P[\alpha|DM]$ 
can be evaluated using Bayes' theorem as
\begin{eqnarray} 
P[\alpha|DM]&=&\int \mathcal{D}\rho\, P[\rho|DM(\alpha)]P[\alpha|M] \\
&\propto& P[\alpha|M]\int\mathcal{D}\rho\,e^{\alpha S[\rho,m]-L[\rho]}\,.
\end{eqnarray} 
Using the saddle-point approximation and the 
Laplace rule ($P[\alpha|M]=\text{const}$), one obtains
\begin{eqnarray} 
P[\alpha|DM]\propto 
\exp\left(\frac{1}{2}\sum_k\text{ln}\frac{\alpha}{\alpha+\lambda_k}+ 
\alpha S[\rho_\alpha,m]-L[\rho_\alpha]\right)\,,\notag 
\end{eqnarray} 
where the $\lambda_k$ are eigenvalues of the following 
real symmetric matrix in functional space 
\begin{eqnarray} 
\Lambda_{ij}=\sqrt{\rho_i}\frac{\partial^2L}{\partial\rho_i\partial\rho_j}
\sqrt{\rho_j}\bigg|_{\rho=\rho_\alpha}\,.
\end{eqnarray} 
Normalizing $P[\alpha|DM]$ and using Eq.\ \eqref{eq:specmem} one 
finally obtains $\rho_{\rm MEM}$.

\section{Numerical Results}\label{sec:numrlt} 
\begin{figure} 
\centering 
\includegraphics[width=0.9\linewidth]{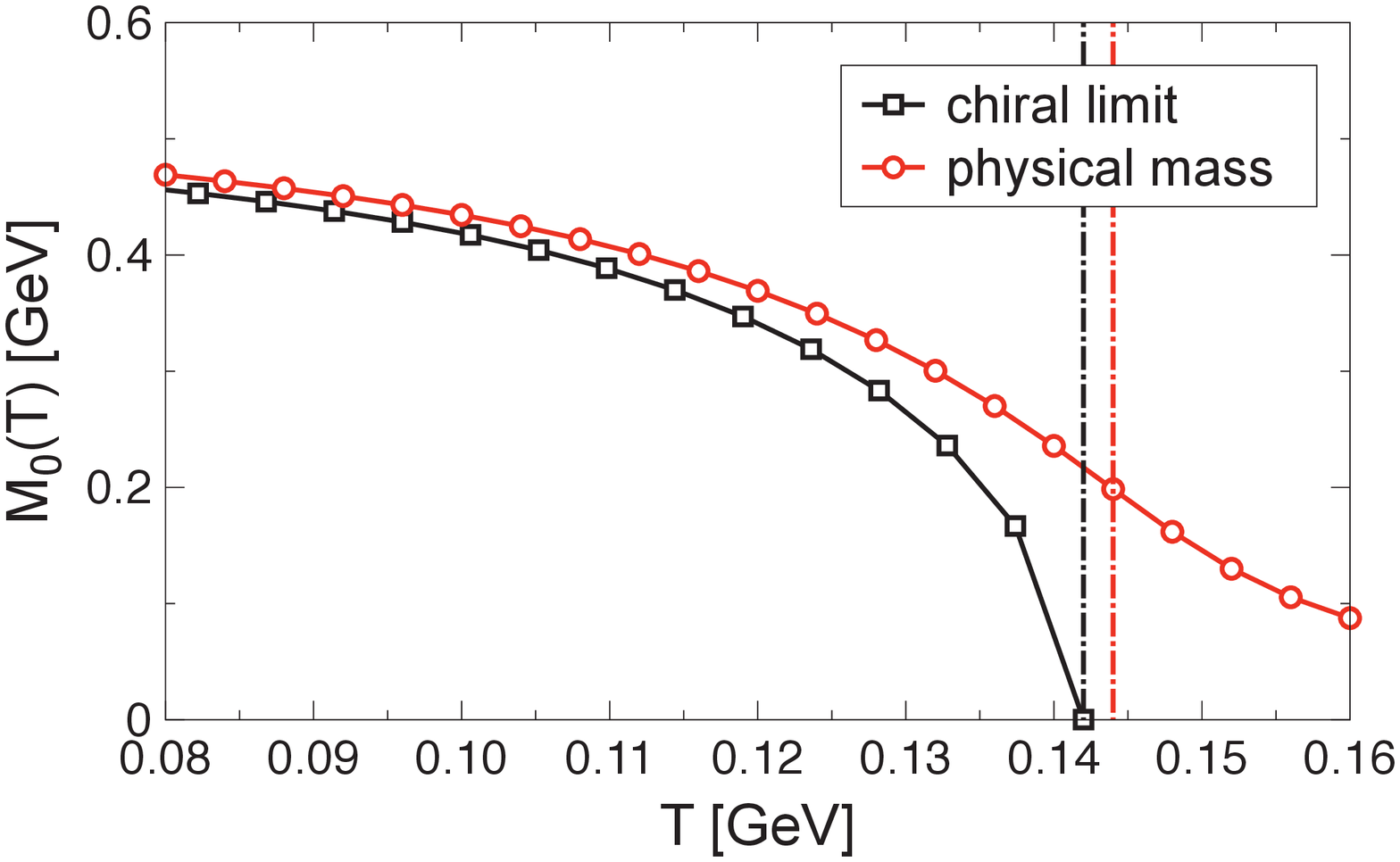} 
\includegraphics[width=0.9\linewidth]{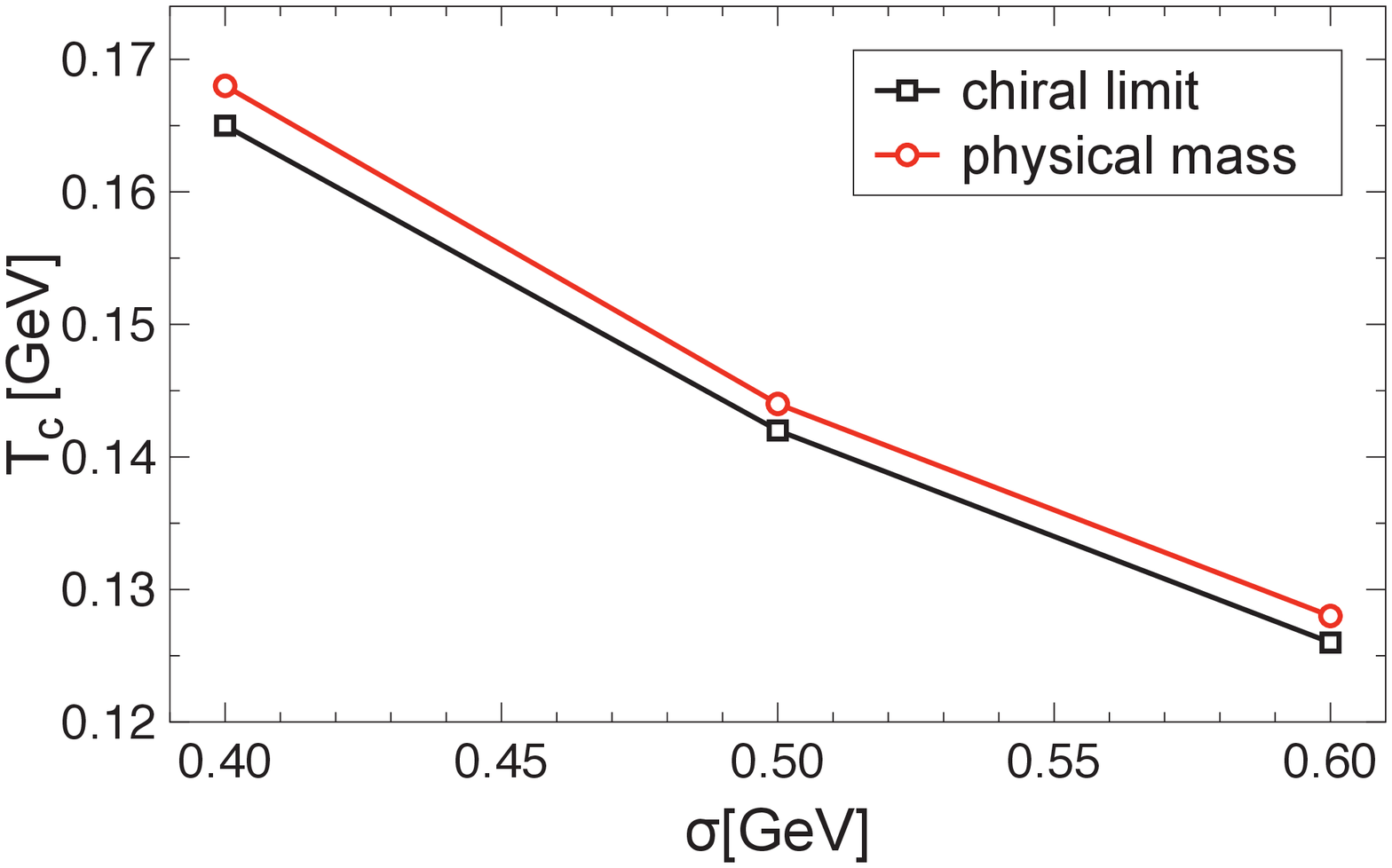} 
\caption{(color online) Upper panel: behavior of the dynamical quark mass 
with temperature (for $\sigma=0.5$\,GeV).
The black line is the chiral limit, the red line is for the physical
value of the current quark mass.
The black dashed line denotes the critical temperature $T_c$ of the 
second-order phase transition in the chiral limit, the red dashed
line denotes the steepest-descent point for the dynamical quark
mass, i.e., the pseudo-critical temperature for the physical current
quark mass. Lower panel: Dependence of the (pseudo-)critical temperature
$T_c$ on the interaction width $\sigma$ in our model.
\label{fig:mT}} 
\end{figure} 
At zero temperature, $T=0$, the largest contribution to the 
constituent quark mass comes from DCSB which dominates 
low-energy hadron physics. With increasing temperature, $T>0$, the 
dynamical quark mass decreases, which indicates a partial restoration 
of chiral symmetry. In the chiral limit, there exists a critical 
temperature $T_c$ where the dynamical quark mass drops to zero and 
chiral symmetry is completely restored through a second-order phase 
transition \cite{Qin:2011ci}. Because of nonzero current quark masses,
chiral symmetry is not exact. Instead of a second-order phase
transition, a crossover happens at some pseudo-critical temperature
$T_{c}$ which is defined by the steepest-descent point for the 
dynamical quark mass. For these two cases, the behavior of the
dynamical quark masses with temperature is illustrated in the
upper panel of Fig.\ \ref{fig:mT}. The (pseudo-)critical temperatures 
have been indicated as vertical dashed lines.

Using our model parameters which are able to provide a 
uniformly good description of vacuum properties of pseudoscalar 
and vector mesons with masses$\,\lesssim 1\,$GeV, we calculate the 
dependence of the (pseudo-)critical temperatures on the interaction 
width $\sigma$, which is shown in the lower panel of Fig.\
\ref{fig:mT}: 
$T_c$ monotonically decreases with increasing $\sigma$. This 
behavior is consistent with results obtained in Ref. \cite{Qin:2011ci}.
Remarkably, the critical temperature range overlaps well with 
that obtained by lattice QCD, i.e., $T_c\in[0.146,0.170]\,$GeV 
\cite{Aoki:2009sc}. 

\begin{figure}
\centering 
\includegraphics[width=0.9\linewidth]{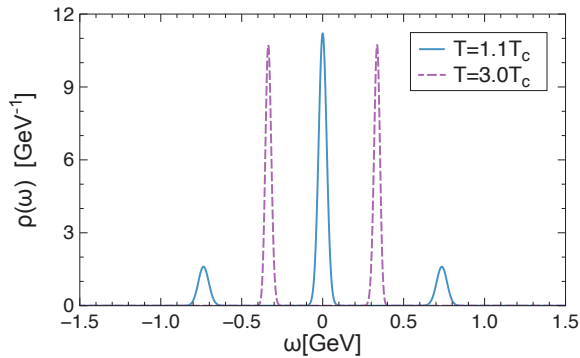} 
\caption{(color online) Typical behavior of the spectral 
function at $T>T_c$ (following Ref.\ \cite{Qin:2011hp}).} \label{fig:specA}
\end{figure}

Above the critical temperature $T_c$, the quark spectral function has 
been studied by both perturbative and nonperturbative approaches. 
At $T>3T_c$ where perturbation theory (hard-thermal-loop resummation) 
works, the properties of the QGP are dominated by two collective 
excitations: thermal and plasmino excitations \cite{LeBellac:2000wh}. 
At $T\gtrsim T_c$, experimental observables indicate that 
nuclear matter is a strongly-coupled QGP (sQGP) \cite{Song:2009ee}. 
In this temperature region, perturbation theory fails while the 
nonperturbative DSE approach predicts 
a novel zero excitation mode in addition to the normal thermal and 
plasmino ones \cite{Qin:2011hp}.
The typical behavior of the quark spectral function is illustrated in 
Fig.\ \ref{fig:specA}. Here, the spectral function is 
positive definite and each peak corresponds to an excitation mode.

\begin{figure}
\centering 
\includegraphics[width=0.9\linewidth]{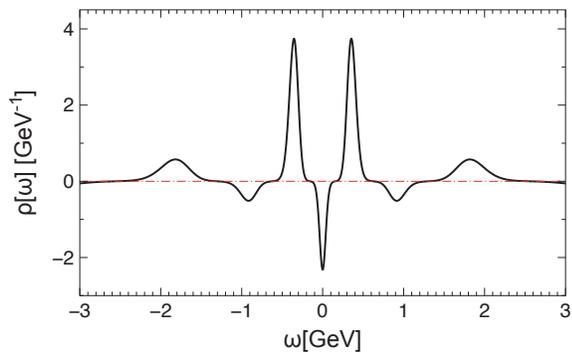} 
\caption{(color online) The behavior of the 
quark spectral function at $T=0.8T_c$ 
($\sigma=0.5\,$GeV, chiral limit).}
\label{fig:specB} 
\end{figure} 
\begin{figure}
\centering 
\includegraphics[width=0.9\linewidth]{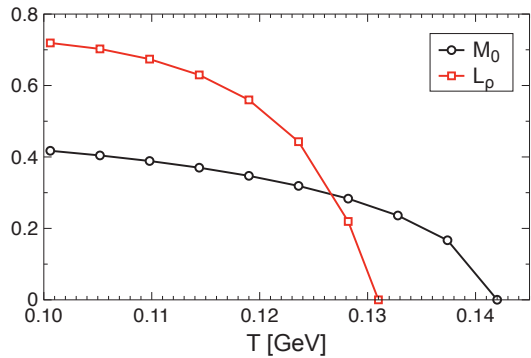}
\caption{(color online) The dynamical quark mass $M_0$ and 
the deconfinement order parameter $L_\rho$ as a function of 
temperature ($\sigma=0.5\,$GeV, chiral limit).} \label{fig:MLT}
\end{figure}
\begin{figure} 
\centering 
\includegraphics[width=0.9\linewidth]{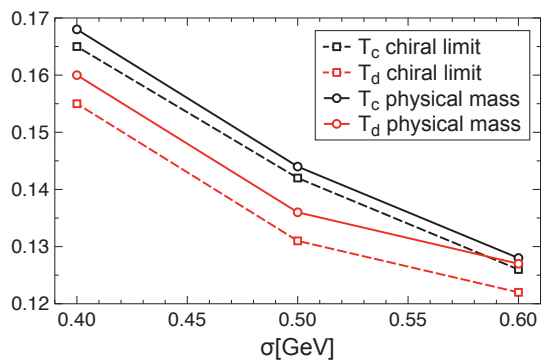} 
\caption{(color online) The (pseudo-)critical temperature $T_c$  (black)
and the deconfinement temperature $T_d$ (red) vs.\
the interaction width $\sigma$ in the chiral limit (dashed lines)
and for a physical current quark mass (full lines).} \label{fig:tcd}
\end{figure}

Below the critical temperature $T_c$, the system is nonperturbative 
because of DCSB and/or confinement. Nevertheless, the spectral 
function computed from the solution of the truncated gap equation 
can provide some nontrivial information about the system. 
We first calculate the quark spectral function at $T=0.8T_c$ with
the interaction width $\sigma=0.5\,$GeV and in the chiral limit, which is 
plotted in Fig.\ \ref{fig:specB}. It is found that the quark spectral 
function exhibits some negative peaks and thus obviously $L_\rho>0$. 
Although the physical meaning of those negative peaks is unclear, 
it still makes sense to analyze how their behavior changes with 
temperature. We found that the structure of the nonpositive 
spectral function remains unchanged while the residues of the
negative peaks, i.e., $L_\rho$, decrease with increasing temperature. 
Notably, there exists a critical temperature $T_d$ where 
$L_\rho$ drops to zero, which signals the positivity restoration of 
the spectral function and deconfinement. The calculated behavior of 
$L_\rho$ is shown in Fig.\ \ref{fig:MLT} in comparison with that of
the dynamical mass $M_0$, which indicates that 
$T_d\lesssim T_c$. Next, we calculate the dependence of $T_d$ 
on the interaction width $\sigma$ both in the chiral limit and with a
physical current quark mass, which is shown in Fig.\ \ref{fig:tcd}: 
$T_d$ monotonically decreases with increasing $\sigma$, and 
$T_d$ is slightly smaller than $T_c$. The difference between 
$T_d$ and $T_c$ for a physical current quark mass is smaller than 
that obtained in the chiral limit. Specifically, 
when $\sigma\in[0.4,0.6]\,$GeV, we have $T_d\sim 94\% T_c$ in the 
chiral limit and $T_d\sim 96\% T_c$ with a physical light quark mass. 
The numerical results are presented in Table \ref{tab:tcd}.
Our results are consistent with Ref.\ \cite{Mueller:2010ah} which also found
positivity violations of the Schwinger function below $T_c$.

\begin{table}
\centering 
\caption{Critical temperatures of chiral symmetry restoration $T_c$ 
and deconfinement $T_d$ for different parameters (dimensionful 
quantities reported in GeV, $\Delta=T_c-T_d$). \label{tab:tcd}}
\begin{tabular*}{1.0\linewidth}{@{\extracolsep{\fill}}c|ccc|ccc}\hline\hline
$\sigma$ & $T_c^0$ & $T_d^0$ & $\Delta^0/T_c^0$ & $T_c^{\rm m}$ &
$T_d^{\rm m}$ &  $\Delta^m/T_c^{\rm m}$ \\\hline
0.4 & 0.165 & 0.155 & $6.1\%$ & 0.168 & 0.160 & $4.8\%$ \\\hline 
0.5 & 0.142 & 0.131 & $7.7\%$ & 0.144 & 0.136 & $5.6\%$ \\\hline 
0.6 & 0.126 & 0.122 & $3.2\%$ & 0.128 & 0.127 & $0.8\%$ \\\hline 
~avg.~ & 0.144 & 0.136 & $5.7\%$ & 0.147 & 0.141 & $3.7\%$
\\\hline\hline  
\end{tabular*} 
\end{table} 

By defining a confinement scale $r_\sigma=1/\sigma$, it is 
apparent that both $T_c$ and $T_d$ increase with increasing
$r_\sigma$, or $T_{c,d} \propto r_\sigma$. Considering that the
difference between $T_c$ and $T_d$ is just several MeVs, while the 
systematic uncertainty introduced by our approximations is certainly
larger, it is reasonable to claim that chiral symmetry restoration and
deconfinement coincide at nonzero temperature and zero chemical potential.

\section{Summary and Remarks} 
At nonzero temperature and zero chemical potential, we computed the 
quark spectral function via the MEM from a solution of
the QCD gap equation. For the latter, we used a rainbow 
interaction kernel which phenomenologically models recent results from 
DSE and lattice QCD. As a criterion for the positivity violation 
and restoration of the quark spectral function, we proposed an
order parameter $L_\rho$ which is directly related to the integral 
of the spectral function's negative part and obviously vanishes for 
positive definite spectral functions. We indeed found that 
$L_\rho>0$ at low temperature while $L_\rho\equiv0$ at high
temperature, i.e., there exists a critical temperature $T_d$ where $L_\rho$ 
drops to zero. Here, the positivity of the quark spectral function
is restored and quarks become asymptotic particles for $T>T_d$. 
Therefore, we conjecture that the deconfinement phase transition 
happens at $T=T_d$. Using our model setup, which can uniformly well describe 
vacuum properties of pseudoscalar and vector mesons with 
masses$\,\lesssim 1\,$GeV, the critical temperature 
of deconfinement comes out slightly smaller than that of chiral symmetry
restoration, i.e., $T_d\lesssim T_c$. Within the systematic
uncertainties of our approach, however, it is 
reasonable to conclude that chiral symmetry restoration 
and deconfinement coincide.

It is generally expected that nuclear matter has a rich phase
structure because of the interplay between DCSB and confinement at 
nonzero chemical potential. It would therefore be interesting to
extend the present study to the case of nonzero chemical potential.
The difference between $T_c$ and $T_d$ could then be more pronounced and
give rise to the so-called quarkyonic phase 
\cite{McLerran:2007dk,McLerran:2009fua}. 

\section*{acknowledgement}
S.-x.\ Qin would like to thank C. S. Fischer and Y.-x.\ Liu for helpful discussions. 
The work of S.-x.\ Qin was supported by the 
Alexander von Humboldt Foundation through a Postdoctoral Research Fellowship.


\begin{thebibliography}{99}

\bibitem{Wilson:1974sk}
K.~G. Wilson,
\newblock Phys. Rev. D {\bf 10}, 2445 (1974).

\bibitem{Eichten:1978dh}
E.~Eichten, K.~Gottfried, T.~Kinoshita, K.~Lane, and T.~Yan,
\newblock Phys. Rev. D {\bf 17}, 3090 (1978).

\bibitem{Bali:2000gf}
G.~Bali,
\newblock Phys. Rept. {\bf 343}, 1 (2001).

\bibitem{Bali:2005bz}
G.~Bali {\em et~al.},
\newblock Phys. Rev. D {\bf 71}, 114513 (2005).

\bibitem{Svetitsky:1982gs}
B.~Svetitsky and L.~G. Yaffe,
\newblock Nucl. Phys. B {\bf 210}, 423 (1982).

\bibitem{Pisarski:2000eq}
R.~D. Pisarski,
\newblock Phys. Rev. D {\bf 62}, 111501 (2000).

\bibitem{Bilgici:2008qy}
E.~Bilgici, F.~Bruckmann, C.~Gattringer, and C.~Hagen,
\newblock Phys. Rev. D {\bf 77}, 094007 (2008).

\bibitem{Fischer:2009wc}
C.~S. Fischer,
\newblock Phys. Rev. Lett. {\bf 103}, 052003 (2009).

\bibitem{Fischer:2009gk}
C.~S. Fischer and J.~A. Mueller,
\newblock Phys. Rev. D {\bf 80}, 074029 (2009).

\bibitem{ROBERTS:1992jd}
C.~D. Roberts,
\newblock Int. J. Mod. Phy. A {\bf 7}, 5607 (1992).

\bibitem{Roberts:1994hs}
C.~D. Roberts and A.~G. Williams,
\newblock Prog. Part. Nucl. Phys. {\bf 33}, 477 (1994).

\bibitem{ROBERTS:2008cq}
C.~D. Roberts,
\newblock Prog. Part. Nucl. Phys. {\bf 61}, 50 (2008).

\bibitem{Maris:1995ns}
P.~Maris,
\newblock Phys. Rev. D {\bf 52}, 6087 (1995).

\bibitem{Bhagwat:2002tx}
M.~Bhagwat, M.~Pichowsky, and P.~C. Tandy,
\newblock Phys. Rev. D {\bf 67}, 054019 (2003).

\bibitem{Yuan:2010iy}
W.~Yuan, S.-x. Qin, H.~Chen, and Y.-x. Liu,
\newblock Phys. Rev. D {\bf 81}, 114022 (2010).

\bibitem{Bryan:1990tva}
R.~K. Bryan,
\newblock Eur. Biophys. J. {\bf 18}, 165 (1990).

\bibitem{Nickel:2007bm}
D.~Nickel,
\newblock Ann. Phys. {\bf 322}, 1949 (2007).

\bibitem{Mueller:2010ah}
J.~A. Mueller, C.~S. Fischer, and D.~Nickel,
\newblock Eur. Phys. J. C {\bf 70}, 1037 (2010).

\bibitem{Asakawa:2000tr}
M.~Asakawa, T.~Hatsuda, and Y.~Nakahara,
\newblock Prog. Part. Nucl. Phys. {\bf 46}, 459 (2001).

\bibitem{Qin:2011kka}
S.-x. Qin, L.~Chang, Y.-x. Liu, C.~Roberts, and D.~Wilson,
\newblock Phys. Rev. C {\bf 84}, 042202 (2011).

\bibitem{Qin:2011xq}
S.-x. Qin, L.~Chang, Y.-x. Liu, C.~D. Roberts, and D.~J. Wilson,
\newblock Phys. Rev. C {\bf 85}, 035202 (2012).

\bibitem{Aguilar:2009ik}
A.~Aguilar, D.~Binosi, J.~Papavassiliou, and J.~Rodr{\'\i}guez-Quintero,
\newblock Phys. Rev. D {\bf 80}, 085018 (2009).

\bibitem{Aguilar:2012rz}
A.~Aguilar, D.~Binosi, and J.~Papavassiliou,
\newblock Phys. Rev. D {\bf 86}, 014032 (2012).

\bibitem{Bowman:2004gt}
P.~O. Bowman, U.~M. Heller, D.~B. Leinweber, M.~B. Parappilly, and A.~G.
  Williams,
\newblock Phys. Rev. D {\bf 70}, 034509 (2004).

\bibitem{Oliveira:2011ds}
O.~Oliveira and P.~Bicudo,
\newblock J. Phys. G: Nucl. Part. Phys. {\bf 38}, 045003 (2011).

\bibitem{Boucaud:2010gr}
Ph.~Boucaud {\em et~al.}, 
\newblock  Phys. Rev. D {\bf 82}, 054007 (2010).

\bibitem{Qin:2011hp}
S.-x. Qin, L.~Chang, Y.-x. Liu, and C.~Roberts,
\newblock Phys. Rev. D {\bf 84}, 014017 (2011).

\bibitem{Bender:1996kg}
A.~Bender, C.~D. Roberts, and L.~v. Smekal,
\newblock Phys. Lett. B {\bf 380}, 7 (1996).

\bibitem{Hobson:1998bz}
M.~Hobson and A.~Lasenby,
\newblock Mon. Not. R. Astron. Soc. {\bf 298}, 905 (1998).

\bibitem{Balandin:1999ab}
A.~Balandin and A.~Kaneko,
\newblock Inverse Problems {\bf 15}, 445 (1999).

\bibitem{Qin:2011ci}
S.-x. Qin, L.~Chang, H.~Chen, Y.-x. Liu, and C.~D. Roberts,
\newblock Phys. Rev. Lett. {\bf 106}, 172301 (2011).

\bibitem{Aoki:2009sc}
Y.~Aoki {\em et~al.},
\newblock JHEP {\bf 0906}, 088 (2009).

\bibitem{LeBellac:2000wh}
M.~Le~Bellac,
\newblock {\em {Thermal Field Theory}} (Cambridge University Press, 2000).

\bibitem{Song:2009ee}
H.~Song and U.~Heinz,
\newblock J. Phys. G: Nucl. Part. Phys. {\bf 36}, 064033 (2009).

\bibitem{McLerran:2007dk}
L.~McLerran and R.~D. Pisarski,
\newblock Nucl. Phys. A {\bf 796}, 83 (2007).

\bibitem{McLerran:2009fua}
L.~McLerran, K.~Redlich, and C.~Sasaki,
\newblock Nucl. Phys. A {\bf 824}, 86 (2009).

\end{thebibliography}
\end{document}